\RequirePackage{lineno}
\documentclass[twocolumn,showpacs,superscriptaddress,amsmath,amssymb]{revtex4-1}
\usepackage[colorlinks,linkcolor=blue,anchorcolor=blue,citecolor=blue,urlcolor=blue]{hyperref}
\usepackage{graphicx}
\usepackage{epsfig}
\usepackage{dcolumn}
\usepackage{bm}
\usepackage{overpic}
\usepackage{color}
\usepackage{multirow}
\usepackage{subfigure}
\usepackage{ulem}

\let\oldequation\equation
\let\oldendequation\endequation

\renewenvironment{equation}
{\linenomathNonumbers\oldequation}
{\oldendequation\endlinenomath}

\begin{document}
\newcommand{\ks}{K_{S}^{0}}
\newcommand{\EP}{e^{+}}
\newcommand{\EM}{e^{-}}
\newcommand{\epm}{e^{\pm}}
\newcommand{\vpho}{\gamma^{\ast}}
\newcommand{\qqbar}{q\bar{q}}

\newcommand{\ee}{e^{+}e^{-}}
\newcommand{\mm}{\mu^{+}\mu^{-}}
\newcommand{\alfs}{\alpha_{s}}
\newcommand{\alfmz}{\alpha(M_{Z}^{2})}
\newcommand{\amu}{a_{\mu}}
\newcommand{\Lam}{\Lambda_{c}}
\newcommand{\lam}{\Lambda_{c}^{+}}
\newcommand{\lambar}{\bar{\Lambda}_{c}^{-}}
\newcommand{\Lambdac}{\Lambda_{c}}
\newcommand{\mbc}{M_{BC}}
\newcommand{\dele}{\Delta E}
\newcommand{\ebm}{E_{\textmd{beam}}}
\newcommand{\ecm}{E_{\textmd{c.m.}}}
\newcommand{\pbm}{p_{\textmd{beam}}}
\newcommand{\MuMu}{\mu\mu}
\newcommand{\mumu}{\mu\mu}
\newcommand{\tata}{\tau^{+}\tau^{-}}
\newcommand{\pipi}{\pi^{+}\pi^{-}}
\newcommand{\kpkm}{K^{+}K^{-}}
\newcommand{\gaga}{\gamma\gamma}
\newcommand{\twopho}{\ee+X}
\newcommand{\sqs}{\sqrt{s}}
\newcommand{\sqsp}{\sqrt{s^{\prime}}}
\newcommand{\da}{\Delta\alpha}
\newcommand{\das}{\Delta\alpha(s)}
\newcommand{\dimu}{\ee \ra \mumu}
\newcommand{\dedx}{\textmd{d}E/\textmd{d}x}
\newcommand{\chip}{\chi_{\textmd{Prob}}}
\newcommand{\chiP}{\chi_{p}}
\newcommand{\evz}{V_{z}^{\textmd{evt}}}
\newcommand{\evzloose}{V_{z,\textmd{loose}}^{\textmd{evt}}}
\newcommand{\avz}{V_{z}^{\textmd{ave}}}
\newcommand{\Ngd}{N_{\textmd{good}}}
\newcommand{\Ncru}{N_{\textmd{crude}}}
\newcommand{\pio}{\pi^{0}}
\newcommand{\rpid}{r_{\textmd{PID}}}

\newcommand{\Nhxobs}{N_{h+X}^{\textmd{obs}}}
\newcommand{\Nhobs}{N_{h}^{\textmd{obs}}}
\newcommand{\Npioxobs}{N_{\pi^{0}+X}^{\textmd{obs}}}
\newcommand{\Nksxobs}{N_{\ks+X}^{\textmd{obs}}}
\newcommand{\Npioobs}{N_{\pi^{0}}^{\textmd{obs}}}
\newcommand{\Nksobs}{N_{\ks}^{\textmd{obs}}}
\newcommand{\Netaobs}{N_{\eta}^{\textmd{obs}}}
\newcommand{\Nhxtru}{N_{h+X}^{\textmd{tru}}}
\newcommand{\Nhtru}{N_{h}^{\textmd{tru}}}
\newcommand{\Npiotru}{N_{\pi^{0}}^{\textmd{tru}}}
\newcommand{\Nkstru}{N_{\ks}^{\textmd{tru}}}
\newcommand{\Nbarhxobs}{\bar{N}_{h+X}^{\textmd{obs}}}
\newcommand{\Nbarhobs}{\bar{N}_{h}^{\textmd{obs}}}
\newcommand{\Nbaretaobs}{\bar{N}_{\eta}^{\textmd{obs}}}
\newcommand{\Nbarpioobs}{\bar{N}_{\pi^{0}}^{\textmd{obs}}}
\newcommand{\Nbarksobs}{\bar{N}_{\ks}^{\textmd{obs}}}
\newcommand{\Nbarhxtru}{\bar{N}_{h+X}^{\textmd{tru}}}
\newcommand{\Nbarhtru}{\bar{N}_{h}^{\textmd{tru}}}
\newcommand{\Nbaretatru}{\bar{N}_{\eta}^{\textmd{tru}}}
\newcommand{\Nbarpiotru}{\bar{N}_{\pi^{0}}^{\textmd{tru}}}
\newcommand{\Nbarkstru}{\bar{N}_{\ks}^{\textmd{tru}}}

\newcommand{\Nhadtot}{N_{\textmd{had}}^{\textmd{tot}}}
\newcommand{\Nhadobs}{N_{\textmd{had}}^{\textmd{obs}}}
\newcommand{\Nbarhadobs}{\bar{N}_{\textmd{had}}^{\textmd{obs}}}
\newcommand{\Nhadtru}{N_{\textmd{had}}^{\textmd{tru}}}
\newcommand{\Nbarhadtru}{\bar{N}_{\textmd{had}}^{\textmd{tru}}}
\newcommand{\Nhadphy}{N_{\textmd{had}}}

\newcommand{\cshadobs}{\sigma_{\textmd{had}}^{\textmd{obs}}}
\newcommand{\effhad}{\vap_{\textmd{had}}}
\newcommand{\efftrg}{\vap_{\textmd{trig}}}
\newcommand{\lint}{\mathcal{L}_{\textmd{int}}}
\newcommand{\Nbkg}{N_{\textmd{bkg}}}
\newcommand{\NbkgTot}{N_{\textrm{bkg}}^{\textrm{Tot}}}
\newcommand{\csbkg}{\sigma_{\textmd{bkg}}}
\newcommand{\Nmcsur}{N_{\textmd{MC}}^{\textmd{sur}}}
\newcommand{\Nmcsurori}{N_{\textmd{MC}}^{\textmd{sur,nom.}}}
\newcommand{\Nmcsurwtd}{N_{\textmd{MC}}^{\textmd{sur,wtd.}}}
\newcommand{\Nmcgen}{N_{\textmd{MC}}^{\textmd{gen}}}
\newcommand{\vap}{\varepsilon}
\newcommand{\chisq}{\chi^{2}}
\newcommand{\cshadphy}{\sigma_{\textmd{had}}^{\textmd{phy}}}
\newcommand{\cshadtot}{\sigma_{\textmd{had}}^{\textmd{tot}}}
\newcommand{\cshadborn}{\sigma_{\textmd{had}}^{0}}
\newcommand{\cshadborncon}{\sigma_{\textmd{con}}^{0}}
\newcommand{\cshadbornres}{\sigma_{\textmd{res}}^{0}}
\newcommand{\csdimuborn}{\sigma_{\mu\mu}^{0}}
\newcommand{\rpqcd}{R_{\textmd{pQCD}}}
\newcommand{\Nprod}{N_{\textmd{prod}}}
\newcommand{\Nhadnet}{N_{\textrm{had}}^{\textrm{net}}}
\newcommand{\Delrel}{\Delta_{\textrm{rel}}}

\newcommand{\fourpionchg}{\pipi\pipi}
\newcommand{\fourpionneu}{\pipi\pi^{0}\pi^{0}}
\newcommand{\sixpionchg}{3(\pipi)}
\newcommand{\thrpionneu}{\pipi\pi^{0}}
\newcommand{\twopionchg}{\pipi}

\newcommand{\Nsurnpion}{N_{\textmd{sur}}^{n\pi}}
\newcommand{\Ngennpion}{N_{\textmd{gen}}^{n\pi}}
\newcommand{\Ngentot}{N_{\textmd{gen}}^{\textmd{tot}}}
\newcommand{\effincnpion}{\vap_{n\pi}^{\textmd{inc}}}
\newcommand{\effincnpionp}{\vap_{n\pi}^{\textmd{inc},\prime}}
\newcommand{\effincnonnpion}{\vap_{\textmd{non}-n\pi}^{\textmd{inc}}}
\newcommand{\effexcnpion}{\vap_{n\pi}^{\textmd{exc}}}
\newcommand{\fracnpion}{f_{n\pi}}
\newcommand{\fracnpionp}{f_{n\pi}^{\prime}}
\newcommand{\fracnonnpion}{f_{\textmd{non}-n\pi}}

\newcommand{\Nsurtwopion}{N_{\textmd{sur}}^{2\pi}}
\newcommand{\Ngentwopion}{N_{\textmd{gen}}^{2\pi}}
\newcommand{\effinctwopion}{\vap_{2\pi}^{\textmd{inc}}}
\newcommand{\effinctwopionp}{\vap_{2\pi}^{\textmd{inc},\prime}}
\newcommand{\effincnontwopion}{\vap_{\textmd{non}-2\pi}^{\textmd{inc}}}
\newcommand{\effexctwopion}{\vap_{2\pi}^{\textmd{exc}}}
\newcommand{\fractwopion}{f_{2\pi}}
\newcommand{\fractwopionp}{f_{2\pi}^{\prime}}
\newcommand{\fracnontwopion}{f_{\textmd{non}-2\pi}}

\newcommand{\Nsurthrpion}{N_{\textmd{sur}}^{3\pi}}
\newcommand{\Ngenthrpion}{N_{\textmd{gen}}^{3\pi}}
\newcommand{\effincthrpion}{\vap_{3\pi}^{\textmd{inc}}}
\newcommand{\effincthrpionp}{\vap_{3\pi}^{\textmd{inc},\prime}}
\newcommand{\effincnonthrpion}{\vap_{\textmd{non}-3\pi}^{\textmd{inc}}}
\newcommand{\effexcthrpion}{\vap_{3\pi}^{\textmd{exc}}}
\newcommand{\fracthrpion}{f_{3\pi}}
\newcommand{\fracthrpionp}{f_{3\pi}^{\prime}}
\newcommand{\fracnonthrpion}{f_{\textmd{non}-3\pi}}

\newcommand{\Nsurfourpion}{N_{\textmd{sur}}^{4\pi}}
\newcommand{\Ngenfourpion}{N_{\textmd{gen}}^{4\pi}}
\newcommand{\effincfourpion}{\vap_{4\pi}^{\textmd{inc}}}
\newcommand{\effincfourpionp}{\vap_{4\pi}^{\textmd{inc},\prime}}
\newcommand{\effincnonfourpion}{\vap_{\textmd{non}-4\pi}^{\textmd{inc}}}
\newcommand{\effexcfourpion}{\vap_{4\pi}^{\textmd{exc}}}
\newcommand{\fracfourpion}{f_{4\pi}}
\newcommand{\fracfourpionp}{f_{4\pi}^{\prime}}
\newcommand{\fracnonfourpion}{f_{\textmd{non}-4\pi}}

\newcommand{\Npionprod}{N_{\textmd{prod}}^{4\pi}}
\newcommand{\Ndatasur}{N_{\textmd{data}}^{\textmd{sur}}}
\newcommand{\Nobspion}{N_{\textmd{obs}}^{4\pi}}
\newcommand{\Nhadprod}{N_{\textmd{prod}}^{\textmd{had}}}
\newcommand{\sigmaobs}{\sigma_{\textmd{obs}}}
\newcommand{\effhadp}{\vap_{\textmd{had}}^{\prime}}

\newcommand{\effpion}{\vap_{4\pi}}
\newcommand{\effexcpion}{\vap_{4\pi}^{\textmd{exc}}}
\newcommand{\effincpion}{\vap_{4\pi}^{\textmd{inc}}}
\newcommand{\effincpionI}{\vap_{4\pi}^{\textmd{inc},1}}
\newcommand{\effincpionII}{\vap_{4\pi}^{\textmd{inc},2}}
\newcommand{\effincpionp}{\vap_{4\pi}^{\textmd{inc},\prime}}
\newcommand{\effincremain}{\vap_{\textmd{non}-n\pi}^{\textmd{inc}}}

\newcommand{\fracpion}{f_{4\pi}}
\newcommand{\fracnonpion}{f_{\textmd{non}-4\pi}}
\newcommand{\fracnonpionp}{f_{\textmd{non}-4\pi}^{\prime}}
\newcommand{\fracpionII}{f_{4\pi}^{2}}
\newcommand{\fracpionp}{f_{4\pi}^{\prime}}
\newcommand{\reladiff}{\Delta_{\textmd{rel}}}

\newcommand{\Nsursixpion}{N_{\textmd{sur}}^{6\pi}}
\newcommand{\Ngensixpion}{N_{\textmd{gen}}^{6\pi}}
\newcommand{\effincsixpion}{\vap_{6\pi}^{\textmd{inc}}}
\newcommand{\fracsixpion}{f_{6\pi}}

\newcommand{\etot}{E_{\textmd{tot}}}
\newcommand{\ptot}{p_{\textmd{tot}}}
\newcommand{\plab}{p_{\textmd{Lab}}}
\newcommand{\mpiOI}{M(\pi^{0}_{1})}
\newcommand{\mpiOII}{M(\pi^{0}_{2})}

\newcommand{\widtheeoi}{\varGamma^{\textmd{ee}}_{0,i}}
\newcommand{\widtheeoj}{\varGamma^{\textmd{ee}}_{0,j}}
\newcommand{\widtheeo}{\varGamma^{\textmd{ee}}_{0}}
\newcommand{\widthee}{\varGamma^{\textmd{ee}}}
\newcommand{\widtheeexpi}{\varGamma^{\textmd{ee}}_{\textmd{exp},i}}
\newcommand{\widtheeexp}{\varGamma^{\textmd{ee}}_{\textmd{exp}}}
\newcommand{\widthtoti}{\varGamma^{\textmd{tot}}_{i}}
\newcommand{\widthtot}{\varGamma^{\textmd{tot}}}

\newcommand{\vpqed}{\Pi_{\textmd{QED}}}
\newcommand{\vpqcd}{\Pi_{\textmd{QCD}}}
\newcommand{\vpcon}{\Pi_{\textmd{con}}}
\newcommand{\vpres}{\Pi_{\textmd{res}}}
\newcommand{\vpo}{\Pi_{0}}
\newcommand{\rcon}{R_{\textmd{con}}}
\newcommand{\rres}{R_{\textmd{res}}}
\newcommand{\rexp}{R_{\textmd{exp}}}

\newcommand{\delvert}{\delta_{\textmd{vert}}}
\newcommand{\delvp}{\delta_{\textmd{vac}}}
\newcommand{\delbrem}{\delta_{\gamma}}
\newcommand{\delobs}{\delta_{\textmd{obs}}}
\newcommand{\radiatorsf}{F_{\textmd{SF}}}
\newcommand{\radiatorfd}{F_{\textmd{FD}}}
\newcommand{\DelFD}{\Delta_{\textmd{FD}}}
\newcommand{\DelFDCal}{\Delta_{\textmd{cal}}}
\newcommand{\DelFDcs}{\Delta_{\sigma}}
\newcommand{\DelFDvp}{\Delta_{\textmd{vp}}}

\newcommand{\costh}{\cos\theta}
\newcommand{\costhIprg}{\cos\theta_{\textmd{1prg}}}
\newcommand{\costhIIprg}{\cos\theta_{\textmd{2prg}}}
\newcommand{\costhIIIprg}{\cos\theta_{\textmd{3prg}}}
\newcommand{\costhIVprg}{\cos\theta_{\textmd{4prg}}}
\newcommand{\costhrestprg}{\cos\theta_{\textmd{restprg}}}
\newcommand{\emce}{E^{\textmd{ctrk.}}_{\textmd{emc}}}
\newcommand{\emceIprg}{E^{\textmd{ctrk.}}_{\textmd{emc,1prg}}}
\newcommand{\emceIIprg}{E^{\textmd{ctrk.}}_{\textmd{emc,2prg}}}
\newcommand{\emceIIIprg}{E^{\textmd{ctrk.}}_{\textmd{emc,3prg}}}
\newcommand{\emceIVprg}{E^{\textmd{ctrk.}}_{\textmd{emc,4prg}}}
\newcommand{\emcerestprg}{E^{\textmd{ctrk.}}_{\textmd{emc,restprg}}}
\newcommand{\isocosth}{\cos\theta_{\textmd{iso}}}
\newcommand{\isocosthIprg}{\cos\theta_{\textmd{iso,1prg}}}
\newcommand{\isocosthIIprg}{\cos\theta_{\textmd{iso,2prg}}}
\newcommand{\isocosthIIIprg}{\cos\theta_{\textmd{iso,3prg}}}
\newcommand{\isocosthIVprg}{\cos\theta_{\textmd{iso,4prg}}}
\newcommand{\isocosthrestprg}{\cos\theta_{\textmd{iso,restprg}}}
\newcommand{\eop}{E/P}
\newcommand{\eopIprg}{E/P_{\textmd{1prg}}}
\newcommand{\eopIIprg}{E/P_{\textmd{2prg}}}
\newcommand{\eopIIIprg}{E/P_{\textmd{3prg}}}
\newcommand{\eopIVprg}{E/P_{\textmd{4prg}}}
\newcommand{\eoprestprg}{E/P_{\textmd{restprg}}}
\newcommand{\nisogam}{N_{\textmd{isogam}}}
\newcommand{\nisogamIprg}{N_{\textmd{isogam,1prg}}}
\newcommand{\nisogamIIprg}{N_{\textmd{isogam,2prg}}}
\newcommand{\nisogamIIIprg}{N_{\textmd{isogam,3prg}}}
\newcommand{\nisogamIVprg}{N_{\textmd{isogam,4prg}}}
\newcommand{\nisogamrestprg}{N_{\textmd{isogam,restprg}}}
\newcommand{\ptrk}{p_{\textmd{ctrk}}}
\newcommand{\pIprg}{p^{\textmd{ctrk}}_{\textmd{1prg}}}
\newcommand{\pIIprg}{p^{\textmd{ctrk}}_{\textmd{2prg}}}
\newcommand{\pIIIprg}{p^{\textmd{ctrk}}_{\textmd{3prg}}}
\newcommand{\pIVprg}{p^{\textmd{ctrk}}_{\textmd{4prg}}}
\newcommand{\prestprg}{p^{\textmd{ctrk}}_{\textmd{restprg}}}
\newcommand{\tote}{E_{\textmd{vis.}}}
\newcommand{\totevte}{E_{\textmd{tot.}}}
\newcommand{\totevteIprg}{E_{\textmd{tot.}}^{\textmd{1prg}}}
\newcommand{\balanceIprg}{\textmd{Balance}}
\newcommand{\ngamma}{N_{\gamma}}
\newcommand{\ngammaIprg}{N_{\gamma,\textmd{1prg}}}
\newcommand{\ngammaIIprg}{N_{\gamma,\textmd{2prg}}}
\newcommand{\ngammaIIIprg}{N_{\gamma,\textmd{3prg}}}
\newcommand{\ngammaIVprg}{N_{\gamma,\textmd{4prg}}}
\newcommand{\ngammarestprg}{N_{\gamma,\textmd{restprg}}}
\newcommand{\ngoodwt}{N_{\textmd{good}}^{\textmd{Wt}}}
\newcommand{\ngood}{N_{\textmd{good}}}
\newcommand{\npiO}{N_{\pi^{0}}}
\newcommand{\npiOIprg}{N_{\pi^{0}}^{\textmd{1prg}}}
\newcommand{\npiOIIprg}{N_{\pi^{0}}^{\textmd{2prg}}}
\newcommand{\npp}{N_{p}}
\newcommand{\npm}{N_{\bar{p}}}
\newcommand{\nkp}{N_{K^{+}}}
\newcommand{\nkm}{N_{K^{-}}}
\newcommand{\npip}{N_{\pi^{+}}}
\newcommand{\npim}{N_{\pi^{-}}}
\newcommand{\ppp}{P(p^{+})}
\newcommand{\ppm}{P(\bar{p}^{-})}
\newcommand{\pkp}{p(K^{+})}
\newcommand{\pkm}{p(K^{-})}
\newcommand{\ppip}{P(\pi^{+})}
\newcommand{\ppim}{P(\pi^{-})}
\newcommand{\ppiO}{P(\pi^{0})}
\newcommand{\mpiO}{M(\pi^{0})}
\newcommand{\peta}{p_{\eta}}
\newcommand{\meta}{M(\eta)}
\newcommand{\mks}{M(K^{0}_{s})}
\newcommand{\pks}{p_{K^{0}_{s}}}
\newcommand{\mphi}{M(\phi)}
\newcommand{\pphi}{p_{\phi}}
\newcommand{\mIIgam}{M(\gamma\gamma)}
\newcommand{\mIIgamIprg}{M(\gamma\gamma)^{\textmd{1prg}}}
\newcommand{\pIIgam}{p_{\gamma\gamma}}
\newcommand{\mlambda}{M(\Lambda)}
\newcommand{\plambda}{p_{\Lambda}}
\newcommand{\mdO}{M(D^{0})}
\newcommand{\pdO}{p_{D^{0}}}
\newcommand{\mdstarO}{M(D^{\ast 0})}
\newcommand{\pdstarO}{p_{D^{\ast 0}}}
\newcommand{\mdp}{M(D^{\pm})}
\newcommand{\pdp}{p_{D^{\pm}}}
\newcommand{\mdstarp}{M(D^{\ast\pm})}
\newcommand{\pdstarp}{p_{D^{\ast\pm}}}
\newcommand{\mds}{M(D_{s}^{\pm})}
\newcommand{\pds}{p_{D_{s}^{\pm}}}
\newcommand{\mdstars}{M(D_{s}^{\ast\pm})}
\newcommand{\pdstars}{p_{D_{s}^{\ast\pm}}}
\newcommand{\Vr}{V_{r}}
\newcommand{\Vz}{V_{z}}

\newcommand{\gev}{\mathrm{GeV}}
\newcommand{\mev}{\mathrm{MeV}}
\newcommand{\mevcc}{\mathrm{MeV}/c^{2}}
\newcommand{\gevc}{\mathrm{GeV}/c}
\newcommand{\gevcc}{\mathrm{GeV}/c^2}

\newcommand{\nchg}{N_{\textmd{chg}}}
\newcommand{\eff}{\vap}

\newcommand{\critecm}{1.780}

\newcommand{\ENERGYAT}{4575.5}
\newcommand{\ENERGYBT}{4575.5}
\newcommand{\ENERGYCT}{4575.5}
\newcommand{\ENERGYDT}{4575.5}
\newcommand{\ksdecay}{\ks\ra\pi^{+}\pi^{-}}
\newcommand{\phidecay}{\phi\ra K^{+}K^{-}}
\newcommand{\piOdecay}{\pi^{0}\ra\gamma\gamma}
\newcommand{\Lambdadecay}{\Lambda\ra p\pi^{-}}
\newcommand{\DOdecay}{D^{0}\ra K^{-}\pi^{+}}
\newcommand{\DStarOdecay}{D^{\ast0}\ra D^{0}\pi^{0}}
\newcommand{\Dpdecay}{D^{+}\ra K^{+}\pi^{+}\pi^{-}}
\newcommand{\DStarpdecay}{D^{\ast+}\ra D^{0}\pi^{+}}
\newcommand{\Dsdecay}{D^{+}_{s}\ra K^{+}K^{-}\pi^{+}}
\newcommand{\DStarsdecay}{D^{\ast+}_{s}\ra D^{+}_{s}\gamma}
\title{\boldmath Measurements of Normalized Differential Cross Sections of Inclusive $\eta$ Production \\in $\ee$ Annihilation at Energy from 2.0000 to 3.6710 $\gev$}

\author{
\begin{small}
\begin{center}
M.~Ablikim$^{1}$, M.~N.~Achasov$^{4,c}$, P.~Adlarson$^{75}$, O.~Afedulidis$^{3}$, X.~C.~Ai$^{80}$, R.~Aliberti$^{35}$, A.~Amoroso$^{74A,74C}$, Q.~An$^{71,58,a}$, D.~Anderle$^{56}$, Y.~Bai$^{57}$
% ...
\\
\vspace{0.2cm}
(BESIII Collaboration)\\
\vspace{0.2cm} {\it
$^{1}$ Institute of High Energy Physics, Beijing 100049, People's Republic of China\\
% ...
$^{3}$ Bochum  Ruhr-University, D-44780 Bochum, Germany\\
$^{4}$ Budker Institute of Nuclear Physics SB RAS (BINP), Novosibirsk 630090, Russia\\
% ...
$^{35}$ Johannes Gutenberg University of Mainz, Johann-Joachim-Becher-Weg 45, D-55099 Mainz, Germany\\
% ...
$^{56}$ South China Normal University, Guangzhou 510006, People's Republic of China\\
$^{57}$ Southeast University, Nanjing 211100, People's Republic of China\\
$^{58}$ State Key Laboratory of Particle Detection and Electronics, Beijing 100049, Hefei 230026, People's Republic of China\\
% ...
$^{71}$ University of Science and Technology of China, Hefei 230026, People's Republic of China\\
% ...
$^{74}$ University of Turin and INFN, (A)University of Turin, I-10125, Turin, Italy; (B)University of Eastern Piedmont, I-15121, Alessandria, Italy; (C)INFN, I-10125, Turin, Italy\\
$^{75}$ Uppsala University, Box 516, SE-75120 Uppsala, Sweden\\
% ...
$^{80}$ Zhengzhou University, Zhengzhou 450001, People's Republic of China\\
\vspace{0.2cm}
$^{a}$ Deceased\\
% ...
$^{c}$ Also at the Novosibirsk State University, Novosibirsk, 630090, Russia\\
% ...
}
\end{center}
\vspace{0.4cm}
\vspace{0.4cm}
\end{small}
}

\noaffiliation{}

\date{\today}

\begin{abstract}
Using data samples collected with the BESIII detector operating at the BEPCII 
storage ring, the cross section of the inclusive process $e^{+}e^{-}\to \eta + X$, 
normalized by the total cross section of $e^{+}e^{-} \to \textmd{hadrons}$, is 
measured at eight center-of-mass energy points from 2.0000~GeV to 3.6710~GeV. 
These are the first measurements with momentum dependence in this energy region.
Our measurement shows a significant discrepancy 
compared to
the existing 
fragmentation functions. 
To address this discrepancy, a new QCD analysis is performed at the next-to-next-to-leading order 
with hadron mass corrections and higher twist effects, which can explain both the 
established high-energy data and our measurements reasonably well.

\end{abstract}

\maketitle

Fragmentation Functions (FFs) describing the hadronization of 
color-carrying partons into color-neutral particles are a key non-perturbative ingredient of 
the Quantum Chromodynamics (QCD) factorized cross-section,
which separate the perturbative hard part of the cross section from the non-perturbative part. 
Gaining precise knowledge of various FFs will help us to understand the mechanism of hadron production, 
hence improving our understanding of the color confinement property of QCD at long distance.
In addition, FFs play indispensable roles in constraining the proton spin configuration and the nuclear parton distribution functions (PDFs), and
probing the transport properties of the hot and dense QCD medium created in relativistic heavy-ion collisions \cite{Metz:2016swz,JET:2013cls,JETSCAPE:2020mzn}. 

Different from PDFs, FFs provide us with other rich ingredients to explore the
non-perturbative aspects of QCD, due to the various hadrons produced in 
fixed target and collider experiments, while PDFs are mainly limited to protons~\cite{Metz:2016swz}. In the state-of-the-art measurements of FFs, most of the efforts are 
devoted to pions and kaons owing to their more 
abundant
production yields~\cite{ParticleDataGroup:2022pth}. Currently, there is still a lack of  measurements of $\eta$ mesons. 
Compared to pions and kaons, $\eta$ mesons are expected to provide additional information about the hadronization process as their wave function contains all light quarks and antiquarks~\cite{Collins:2023cuo,Gao:2024nkz,Aidala:2010bn}.
In addition, due to the universality property of FFs, one can relate the $\eta$ production in $\ee$ collisions, $pp$ collisions, and semi-inclusive 
deep inelastic scatterings (SIDIS)~\cite{Collins:1989gx,CTEQ:1993hwr}. 
The QCD analysis has shown good agreement with $\ee$ and proton-proton collisions so far by including the only available 
$\ee$ data at a center-of-mass (c.m.) energy ($\sqs$) above $9~\gev$~\cite{Aidala:2010bn}. 
This work tests the validity of a factorized QCD framework at low energy region.

It is well known that the single inclusive $\ee$ annihilation, $\ee\to h + X$, where $h$ is an identified hadron under investigation and $X$ represents 
everything else, provides an effective way to study collinear FFs~\cite{ParticleDataGroup:2022pth}. A widely measured experimental observable is 
\begin{equation} \frac{1}{\sigma(\ee
    \rightarrow \textmd{hadrons})} \frac{\textmd{d}\sigma(\ee
    \rightarrow h + X)}{\textmd{d}
    p_h}, 
\label{equation_1} 
\end{equation} 
where $\sigma(\ee \rightarrow \textmd{hadrons})$ represents the total cross section for $\ee$ annihilation to all possible hadronic final states (referred to as inclusive hadronic events 
hereafter), and $p_h$ denotes the momentum of the identified hadron $h$.
The observable can be interpreted, in terms of the leading order of $\alpha_s$, as $\sum_q e_q^2[ D_q^h(z,\sqrt{s})+D_{\bar{q}}^h(z,\sqrt{s})]$, where 
$e_{q}$ is the fractional charge of the quark $q$, and $D^h_{q/\bar{q}}(z,\sqrt{s})$ is the FF of quark $q$ or antiquark $\bar{q}$ at c.m. energy $\sqrt{s}$.
The variable $z\equiv 2\sqrt{p_h^2c^2+M^2_hc^4}/\sqrt{s}$ denotes the relative energy of the produced hadron $h$ with mass $M_h$.

This Letter,
for the first time,
reports a measurement of the process $\ee\to\eta +X$ at eight c.m.~energy points from $2.0000$ to $3.6710~\gev$, with a $z$ coverage from 0.3 to 0.9.
Such a special energy coverage is expected to be very sensitive to the initial parametrization of FFs, and can test the convergence of fixed order perturbative QCD (pQCD) calculations based on leading twist factorization involving final state hadron production~\cite{Deur:2016tte,Liu:2019srj,Berger:1980qg}. 
In addition, a comprehensive QCD analysis is performed by involving both experimental measurements and theoretical calculations of $\eta$ production in $\ee$ annihilation.
In the analysis, the data reported in this Letter are implemented into the global study of $\eta$ meson FFs, after considering the highest precision pQCD calculation at next-to-next-to-leading order (NNLO) \cite{Mitov:2006wy}, the mass corrections \cite{Accardi:2014qda,MoosaviNejad:2015lgp}, and the possible contributions from higher twist \cite{Liu:2019srj,Berger:1980qg}.
The analysis not only indicates the importance of the BESIII measurement in exploring the $\eta$ meson hadronization mechanism, but also serves as a testing ground for the validity of QCD factorization at leading twist for inclusive hadron production and their associated fixed order pQCD calculations.

The data sets used in this Letter were collected with the BESIII
detector~\cite{BESIII:2009fln} running at BEPCII~\cite{Yu:2016cof}.
Experimentally, the normalized differential cross section characterizing the inclusive
production of identified hadron $h$, as described in Eq.~\eqref{equation_1}, can be
determined with
\begin{equation}
    \frac{\Nhobs}{\Nhadobs}\frac{1}{\Delta p_h}f_{h},
    \label{equation_2}
\end{equation}
where $\Nhadobs$ represents the number of observed hadronic events in the
$\ee$ annihilation at a given c.m.~energy, and $\Nhobs$ denotes the number of 
$\ee \to h + X$ events within a specific momentum range $\Delta p_h$.
The factor $f_{h}$, described later in detail, is a correction factor 
accounting for the global detection efficiency and the initial state 
radiation (ISR) effects. 
Since both $\Nhadobs$ and $\Nhobs$ are obtained from the same data 
sample, the integrated luminosity used in the cross section 
measurement cancels.

For the first step of this analysis, the hadronic events are identified 
using the same selection criteria as described in Ref.~\cite{BESIII:2021wib}.
The Bhabha and $\ee\to\gaga$ events are removed by applying 
dedicated requirements on the electromagnetic calorimeter (EMC) shower information.
Subsequently, a series of criteria are applied to select good charged tracks.
Events with less than two good charged tracks are removed to suppress background processes.
For events with two or three good charged tracks, further requirements 
are employed to suppress the QED-related backgrounds.
Events with more than three good charged tracks are regarded as hadronic 
events directly. Details of the selection of the inclusive hadronic events 
can be found in Ref.~\cite{BESIII:2021wib}. 

Despite comprehensive selection criteria being applied to identify the inclusive hadronic events, there are still residual background events in the data. 
The numbers of the QED-related backgrounds are estimated by analyzing the corresponding Monte Carlo (MC) simulation samples. 
The {\sc geant4}-based~\cite{GEANT4:2002zbu} programs are used to produce these MC samples, where the geometric description of the BESIII detector~\cite{Huang:2022wuo} 
and the interaction between secondary particles and the detector material are included. 
The {\sc babayaga3.5}~\cite{CarloniCalame:2000pz} package is used to generate the $\ee \to \ee$, $\mu^{+}\mu^{-}$, and $\gaga$ processes, while {\sc kkmc}~\cite{Jadach:1999vf} and {\sc evtgen}~\cite{Lange:2001uf} are utilized for the $\ee\to\tau^{+}\tau^{-}$ process. 
The two-photon processes are simulated by dedicated MC generators~\cite{BESIII:2021wib}. 
The beam-associated background events are estimated using a sideband method~\cite{BESIII:2021wib}.
Table~\ref{tab:had} summarizes the integrated luminosities, the number of total selected hadronic events ($\Nhadtot$) and the total remaining backgrounds ($\Nbkg$) at each c.m.~energy, where $\Nhadobs=\Nhadtot-\Nbkg$.

\setlength{\tabcolsep}{5pt}
\begin{table}[!htbp]
\setlength{\abovecaptionskip}{5pt}
\setlength{\belowcaptionskip}{10pt}
\centering
\caption{The integrated luminosities, the numbers of total selected hadronic and residual background events at eight c.m. energy points.}
  \begin{tabular}{c r r l r l}
  \hline
  \hline
  $\sqrt{s}$~(GeV) & $\mathcal{L}$~($\rm pb^{-1}$) & \multicolumn{2}{c}{$\Nhadtot$}\hspace{-5pt} & \multicolumn{2}{c}{$\Nbkg$}\hspace{-5pt}\\
  \hline
 2.0000  &   10.074  &   350298 $\pm$ &\hspace{-10pt} 591   &   8722  $\pm$ &\hspace{-10pt}  93 \\
 2.2000  &   13.699  &   445019 $\pm$ &\hspace{-10pt} 666   &  10737  $\pm$ &\hspace{-10pt} 103 \\
 2.3960  &   66.869  &  1869906 $\pm$ &\hspace{-10pt} 1365  &  47550  $\pm$ &\hspace{-10pt} 218 \\
 2.6444  &   33.722  &   817528 $\pm$ &\hspace{-10pt} 902   &  21042  $\pm$ &\hspace{-10pt} 145 \\
 2.9000  &  105.253  &  2197328 $\pm$ &\hspace{-10pt} 1478  &  56841  $\pm$ &\hspace{-10pt} 238 \\
 3.0500  &   14.893  &   283822 $\pm$ &\hspace{-10pt} 531   &   7719  $\pm$ &\hspace{-10pt}  87 \\
 3.5000  &    3.633  &    62670 $\pm$ &\hspace{-10pt} 249   &   1691  $\pm$ &\hspace{-10pt}  41 \\
 3.6710  &    4.628  &    75253 $\pm$ &\hspace{-10pt} 273   &   6461  $\pm$ &\hspace{-10pt}  80 \\  
  \hline
  \hline
\end{tabular}
\label{tab:had}
\end{table}

From the selected inclusive hadronic events, the $\eta$ candidates are reconstructed via the $\eta \to \gaga$ decay.
Photons are required to have a deposited energy in the EMC of more than $25~\mev$ in the barrel region ($|\costh|<0.80$), and more than $50~\mev$ in the end cap region ($0.86<|\costh|<0.92$), where $\theta$ is defined with respect to the $z$-axis, which is the symmetry axis of the multi-layer drift chamber. 
To exclude showers that originate from charged tracks, the angle subtended by the EMC shower and the position of the closest charged track at the EMC must be greater than 10 degrees as measured from the interaction point.
To suppress electronic noise and showers unrelated to the event, the difference between the EMC time and the event start time is required to be within [0, 700]\,ns.
Since the yield of $\pio$ mesons is much higher comparing to $\eta$ in the inclusive hadronic events~\cite{BESIII:2022zit}, the photons originating from $\pio$ are excluded when reconstructing $\eta$. 
If a pair of photons in one event has an invariant mass within $(115,~155)~\mev/c^2$, which covers five times the $\pio$ mass resolution, both photons in that pair are discarded. 
A test using the control sample $J/\psi\to\kpkm\pipi\eta$ indicates that the exclusion of the $\pio$ photons does not introduce any bias to the invariant-mass distribution of $\eta$. 
Moreover, a check using the MC sample of the inclusive hadronic process demonstrates that the exclusion does not induce any peaking background in the $\eta$ invariant-mass distribution. 
The remaining photons after veto of $\pi^0$ are paired to reconstruct the $\eta$ candidates. 
To suppress background due to photon mis-combinations, the helicity variable of the $\eta$ candidate, defined as $|E_{\gamma1}-E_{\gamma2}|/p_{\gaga}$ where $E_{\gamma1,2}$ represent the deposited energies of photons and $p_{\gaga}$ denotes the momentum of $\eta$ candidate, is required to be less than 0.8.
In this analysis, the inclusive hadronic events are simulated with the {\sc luarlw}
generator~\cite{Andersson:1997xwk,Andersson:1999ui,BESIII:2021wib}, in which among others the signal processes $\ee\to\eta+X$ is contained.

\begin{figure}[!htbp]
\begin{center}
\begin{overpic}[width=0.46\textwidth]{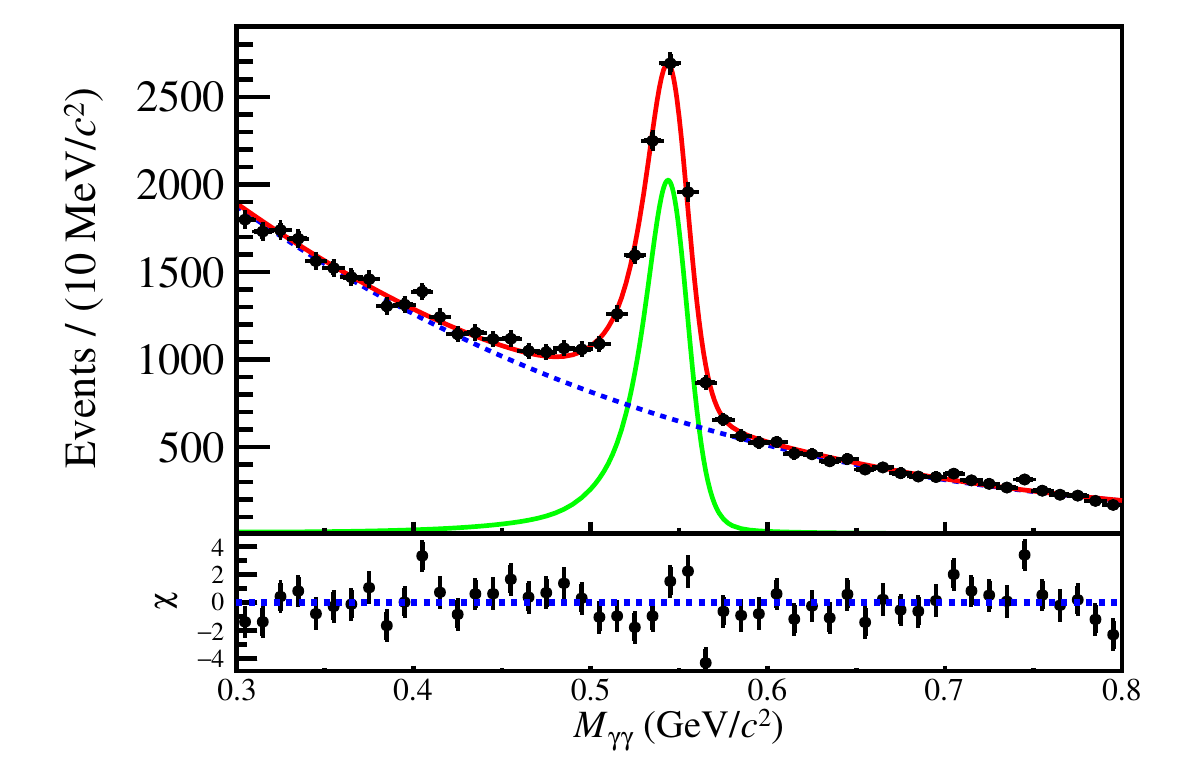}
\end{overpic}
\end{center}
\caption{The $M(\gaga)$ distributions for $\eta$ candidates, with $p_{\gaga}\in(0.4, 0.5)~\gevc$ at $\sqs=2.9000~\gev$. The fit results are overlaid. The black points with error bars are data. The red solid curve is the sum of fit functions, while the green solid and blue dashed curves represent the signal and background, respectively. The pull variable $\chi$, defined as the residual between data and total fit function, normalized by the uncertainty of the data, is shown on the bottom of the figure.}
\label{fig:etafit}
\end{figure}

The $\eta$ candidates are divided into different momentum intervals with the bin width $\Delta \peta=0.1~\gev/c$, approximately 6 times the momentum resolution of $\eta$.
The yield of $\eta$ mesons in each momentum bin is determined by an unbinned maximum likelihood fit.
In the fit, the signal is described by the signal shape of $\eta$, which is extracted from the {\sc luarlw} MC sample, convolved with a Gaussian function corresponding to the difference between data and MC. 
The background is modeled by a second-order polynomial, except for a few momentum bins where a third-order polynomial is used due to a higher mis-combination background level.
Figure~\ref{fig:etafit} shows the fit result of $\eta$ candidates with $p_{\gaga}\in(0.4, 0.5)~\gevc$ at $\sqs=2.9000~\gev$, 
where the higher background at the low mass region is caused by the mis-combinations involving the low energy photons.
To extract the signal shape of $\eta$ from the {\sc luarlw} MC sample, the truth-level $\eta$ mesons decaying to two photons are matched to the reconstructed $\eta$ candidates according to the momentum direction. 
The reconstructed $\eta$ candidate which has the closest momentum direction relative to the truth-level $\eta$ is regarded as matched. 
In addition, the angle of the momenta between truth-level $\eta$ and its matched $\eta$ candidate is required to be less than $25$ degrees. 
The matched $\eta$ candidates make up the signal sample of $\eta$ and their invariant-mass distributions are used as the signal shapes in the fitting procedure.
The obtained $\Netaobs$ in each momentum range of $\eta$ at various c.m.~energies are summarized in the Supplemental Material~\cite{supp}.

The correction factor $f_{h}$,  which scales the observable quantity $\Nhobs/\Nhadobs$ to determine the observable given in Eq.~(\ref{equation_1}) in each momentum bin, 
is extracted from the inclusive hadronic MC sample and is expressed as:
\begin{equation}
    f_{h} = \frac{\Nbarhtru\textmd{(off)}}{\Nbarhadtru\textmd{(off)}}\bigg/\frac{\Nbarhobs\textmd{(on)}}{\Nbarhadobs\textmd{(on)}}.
    \label{equation_3}
\end{equation}
Here, $\bar{N}$ denotes the number of events determined from the inclusive hadronic MC sample, either at observable level, similar to the experimental data, with superscript ``obs'' or 
at truth level with superscript ``tru''.
The terms ``on'' and ``off'' in the parentheses indicate that the corresponding quantities are extracted from the inclusive hadronic MC sample with or without simulating the ISR process, respectively.
In this Letter, $\Nbaretaobs\textmd{(on)}$ is determined with a similar fit to the $M(\gaga)$ distribution of the $\eta$ candidates selected from the inclusive hadronic MC sample.

Extensive comparisons between the {\sc luarlw} generated MC events and the experimental data show that the {\sc luarlw} model can reasonably reproduce the multiplicity and 
kinematic quantity of the $\eta$ mesons~\cite{supp}. 
In addition, good agreements are observed in terms of the invariant mass spectra of $\eta$ in different momentum bins between experimental data and the inclusive hadronic MC sample. 
Thus, the correction applied in this analysis is valid.
The calculated results of $f_{\eta}$ in the different $\eta$ momentum ranges are presented in the Supplemental Material~\cite{supp}.

The systematic uncertainties of the normalized differential cross section are mainly caused by the residual deviations between the signal MC and data samples, the reconstruction efficiency of the $\eta$ candidates, the fit scheme of the $M(\gaga)$ spectrum, and the simulation model of the inclusive hadronic events. 

The approach described in Ref.~\cite{BESIII:2022zit} is applied here to estimate the uncertainty caused by the imperfect simulation of signal events.
Systematic uncertainties of the differential cross section introduced by the determination of $\Nbkg$ are found to be negligible.

For the uncertainty of reconstructing the $\eta$ candidates, several factors are considered, including the identification of photons,  the exclusion of the $\pio$ photons, and the helicity requirement. 
The uncertainty in the photon identification is estimated to be 1\% per photon~\cite{BESIII:2018rdg}, resulting in 2\% uncertainty for each $\eta$ meson.
The uncertainty due to the exclusion of the $\pio$ photons is evaluated by varying the nominal invariant-mass range of $\pio$ to $(111, 159)~\mev/c^2$. 
The differences of the differential cross section are found to be negligible (less than 1\%).
To estimate the uncertainty due to the $\eta$ helicity requirement, the $\eta$ helicity distributions of the $J/\psi\to\kpkm\pipi\eta$ events are compared between data and MC simulation. 
The average relative difference of 2\% is taken as the uncertainty.

To evaluate the uncertainty of the fit scheme, different signal and background description functions are applied to fit the $M(\gaga)$ spectrum. 
For the signal, the Crystal Ball function~\cite{Skwarnicki:1986xj} is used as an alternative model.
Moreover, the different requirements of the match angle, which are $20$ and $30$ degrees, are utilized to extract the alternative signal shapes from the signal MC sample.
For the background, the alternative models are obtained by varying the order of the Chebychev polynomial. 
All the resulting relative differences in the differential cross sections are combined in quadrature, and taken as the systematic uncertainty.

\begin{figure*}[!htbp]
\setlength{\abovecaptionskip}{-0.2cm}
\setlength{\belowcaptionskip}{-0.2cm}
\begin{center}
\begin{overpic}[width=0.9\textwidth]{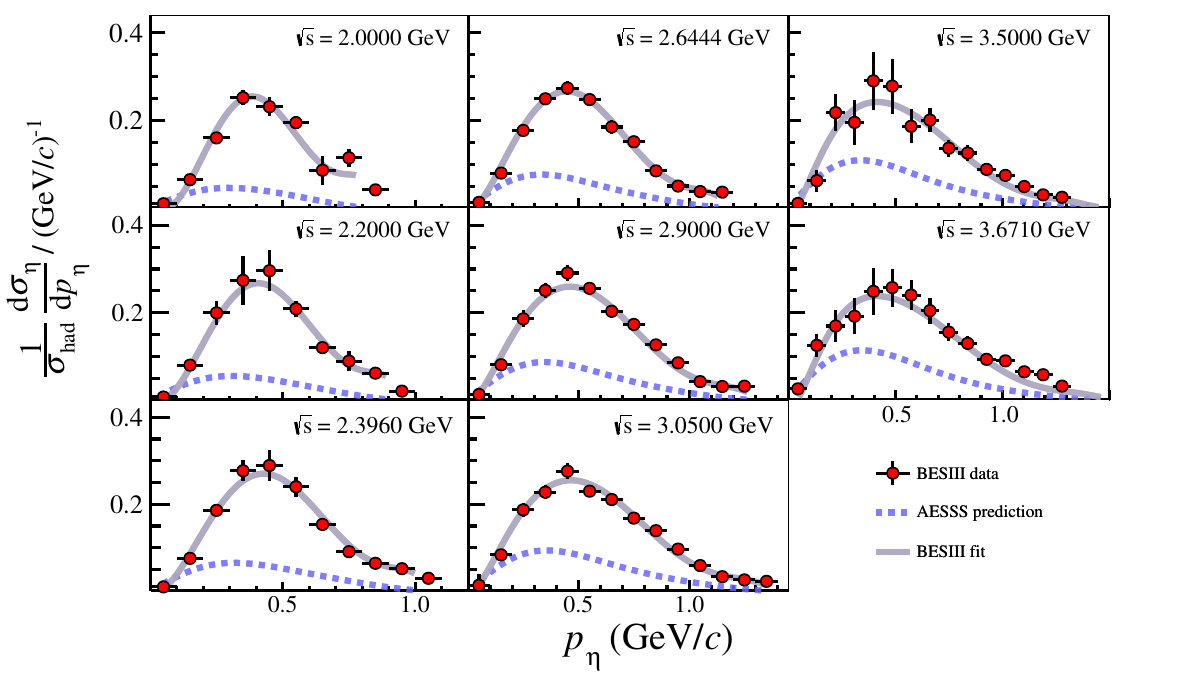}
\end{overpic}
\end{center}
\caption{Normalized differential cross sections of the $\ee\to\eta+X$ process. 
The points with error bars are the measured values. 
The blue dotted curves denote the predictions by using the AESSS FFs, while the curves in gray denote the calculations by using the newly extracted FFs from our fit based on the available $\eta$ production data in $\ee$ annihilation and our BESIII data.
Notice that an upper cut of $z<0.95$, where theoretical curves stop, is employed in our global analysis to avoid large enhancement from threshold logarithms $\propto \log(1-z)$.
The normalized differential cross section in terms of $z$ is shown in the Supplemental Material~\cite{supp}. 
}\label{fig:eta_data_theory}
\end{figure*}

In this Letter, the dominant systematic uncertainty is introduced by the MC simulation model of the inclusive hadronic events. 
According to Eq.~\eqref{equation_3}, the generation fractions of the exclusive processes containing the $\eta$ mesons, which make up the inclusive process $\ee\to\eta+X$, directly affect the correction factors $f_{\eta}$.
To address the corresponding uncertainty, the {\sc hybrid} generator, which was developed in Ref.~\cite{Ping:2016pms} and improved in Ref.~\cite{BESIII:2021wib}, is used as an alternative model to reproduce the inclusive hadronic events. 
The discrepancies observed in the correction factors $f_{\eta}$ relative to the nominal ones are regarded as systematic uncertainties. 

All these individual systematic uncertainties are regarded as uncorrelated with each other therefore are summed in quadrature. The normalized differential cross sections for the inclusive $\eta$ production in $\ee$ annihilation at the eight c.m.~energy points are tabulated in the Supplemental Material~\cite{supp} and shown in Fig.~\ref{fig:eta_data_theory}. 

The blue dotted curve in Fig.~\ref{fig:eta_data_theory} represents a theoretical prediction performed using the $\eta$ FF from the Aidala-Ellinghaus-Sassot-Seele-Stratmann (AESSS) parametrization at the next-to-leading order~\cite{Aidala:2010bn}. 
The AESSS FFs are extracted using data of $\eta$ production in $\ee$ annihilation with $\sqrt{s}\eqsim$ 10, 30, and 90 GeV, and $pp$ collisions with $\sqrt{s}\eqsim$ 200 GeV whose energy scales are higher than the typical BESIII c.m.~energies. 
The AESSS study is based on the well-established de Florian-Sassot-Stratmann (DSS) framework~\cite{deFlorian:2007ekg,deFlorian:2007aj} for FF extractions. 
It consists of a pure next-to-leading order analysis, based on leading-twist pQCD factorization theorems for $\ee$ annihilation and $pp$ collision processes, where the mass of the $\eta$-meson is considered to be negligible and set to be zero. 
The fit in Ref.~\cite{Aidala:2010bn} reveals good agreement among data sets taken at different energy scales.
However, according to Fig.~\ref{fig:eta_data_theory}, the AESSS fit cannot describe the previous data and the BESIII data at the same time.

The gray line in Fig.~\ref{fig:eta_data_theory} shows the calculation with a new extraction of $\eta$ FFs~\cite{supp,Li:2024etc} based on the available $\eta$ production data in $\ee$ annihilation experiments, namely the data sets included in Ref.~\cite{Aidala:2010bn}, except for the un-published BaBar data, and the BESIII data presented in this Letter. 
The ratio of $\chi^2/N_{d.o.f.}$ for this fit is 1.52~\cite{supp,Li:2024etc} which is comparable to that of the AESSS fit (1.91~\cite{Aidala:2010bn}) where only the existing $\ee$ annihilation data are considered. 
For the first time, data at $\sqrt{s}<5~\gev$ are included in such a QCD-based analysis where the analysis framework is extended to NNLO accuracy and the hadron mass corrections and higher twist contributions are considered~\cite{supp,Li:2024etc}.
The inclusion of BESIII data in the original AESSS framework, namely a refit, leads to a significantly large $\chi^2/N_{d.o.f.}$ (12.79) which confirms the disagreement between the AESSS fit and the BESIII data.
Each of the three major effects considered in the new fit plays a fundamental role in achieving the good agreement as shown in Fig.~\ref{fig:eta_data_theory}. 
Hadron mass corrections are well known to be an important effect in the fit of FFs of heavier hadron species, e.g., see discussions in Refs.~\cite{Albino:2008fy,Albino:2005gd,Accardi:2014qda}. 
The extension to NNLO has an important effect on the shape of the observable in the lower-$z$ region, as one can for example explicitly see in the case of the NNLO parton-to-pion FF fit for $\ee$ production in Fig.~4 of Ref.~\cite{Anderle:2016czy}. 
That analysis highlighted the importance of the higher accuracy framework in order to perform a reasonable fit including Belle and BaBar data sets which have c.m.~energy around 10.5~GeV. 
At last, higher twist effects are commonly introduced in PDF analysis when incorporating low energy data, such as JLab data in Ref.~\cite{Owens:2012bv}. 
They are taken into account as extra fit parameters that parameterize an additional $1/Q^2$ dependence to the leading twist expression of the observable. 
In this analysis, both $1/Q^2$ and $1/Q^4$ dependence have to be introduced in order to obtain a good fit. 
The analysis is summarized in Supplemental Material~\cite{supp} and detailed in Ref.~\cite{Li:2024etc}.

In summary, we have measured the normalized differential cross sections of the $\ee\to\eta + X$ processes, using data samples collected from $\sqs=2.0000$ to $3.6710~\gev$.  
The results obtained in this work fill this particular energy region where no such kind of measurements have been reported before. 
A QCD-based analysis shows that in order to explain both high and low-energy data, one needs to consider higher-order contributions as well as higher twist effects. 
It would be interesting to check if a more flexible approach, such as NNFF~\cite{Bertone:2017tyb} or MAPFF~\cite{AbdulKhalek:2022laj}, could describe the data over the full energy range.
These new results in the relatively low energy region provide special ingredients for FF studies, moreover, they will help to enhance our understanding of the QCD factorization theorem at the leading twist and beyond.

The BESIII Collaboration thanks the staff of BEPCII, the IHEP computing center and the supercomputing center of USTC for their strong support. This work is supported in part by National Key R\&D Program of China under Contracts Nos. 2020YFA0406400, 2020YFA0406300; National Natural Science Foundation of China (NSFC) under Contracts Nos. 11635010, 11735014, 11835012, 11935015, 11935016, 11935018, 11961141012, 12025502, 12035009, 12035013, 12061131003, 12192260, 12192261, 12192262, 12192263, 12192264, 12192265, 12221005, 12225509, 12235017, 12122509, 12105276, 11625523, 12205255, 12022512, 12035007, 12150410312; China Postdoctoral Science Foundation under Contracts No. 2019M662152 and No. 2020T130636; the Fundamental Research Funds for the Central Universities, University of Science and
Technology of China under Contract No. WK2030000053; Guangdong Major Project of Basic and Applied
Basic Research No. 2020B0301030008; the Chinese Academy of Sciences (CAS) Large-Scale Scientific Facility Program; Joint Large-Scale Scientific Facility Funds of the NSFC and CAS under Contract Nos. U1832207, U1732263, U1832103, U2032111, U2032105; CAS Key Research Program of Frontier Sciences under Contracts Nos. QYZDJ-SSW-SLH003, QYZDJ-SSW-SLH040; the CAS Center for Excellence in Particle Physics (CCEPP); 100 Talents Program of CAS; The Institute of Nuclear and Particle Physics (INPAC) and Shanghai Key Laboratory for Particle Physics and Cosmology; European Union's Horizon 2020 research and innovation programme under Marie Sklodowska-Curie grant agreement under Contract No. 894790; German Research Foundation DFG under Contracts Nos. 455635585, Collaborative Research Center CRC 1044, FOR5327, GRK 2149; Istituto Nazionale di Fisica Nucleare, Italy; Ministry of Development of Turkey under Contract No. DPT2006K-120470; National Research Foundation of Korea under Contract No. NRF-2022R1A2C1092335; National Science and Technology fund of Mongolia; National Science Research and Innovation Fund (NSRF) via the Program Management Unit for Human Resources \& Institutional Development, Research and Innovation of Thailand under Contract No. B16F640076; Polish National Science Centre under Contract No. 2019/35/O/ST2/02907; The Swedish Research Council; U. S. Department of Energy under Contract No. DE-FG02-05ER41374.

\end{document}